# Interpreting the dependence of mutation rates on age and time

Ziyue Gao[1,+,*], Minyoung J. Wyman[2], Guy Sella[2] and Molly Przeworski[2,3,+]

[1] Committee on Genetics, Genomics and Systems Biology, University of Chicago

[2] Dept. of Biological Sciences, Columbia University

[3] Dept. of Systems Biology, Columbia University

[+] To whom correspondence should be addressed: ziyuegao@uchicago.edu or mp3284@columbia.edu

[*] Current address: 606D Fairchild Center, M.C. 2424, New York, NY 10027

**Abstract**

Mutations can arise from the chance misincorporation of nucleotides during DNA replication or from DNA lesions that are not repaired correctly. We introduce a model that relates the source of mutations to their accumulation with cell divisions, providing a framework for understanding how mutation rates depend on sex, age and absolute time. We show that the accrual of mutations should track cell divisions not only when mutations are replicative in origin but also when they are non-replicative and repaired efficiently. One implication is that the higher incidence of cancer in rapidly renewing tissues, an observation ascribed to replication errors, could instead reflect exogenous or endogenous mutagens. We further find that only mutations that arise from inefficiently repaired lesions will accrue according to absolute time; thus, in the absence of selection on mutation rates, the phylogenetic "molecular clock" should not be expected to run steadily across species.

**Introduction**

Because mutations are the ultimate source of all genetic variation, deleterious and advantageous, mutagenesis has been of central interest even before the discovery of DNA as the genetic material (e.g., Muller, 1927) and developing a model of mutations along the genome is a major focus of current disease mapping studies (Lawrence et al., 2013, Samocha et al., 2014). From many decades of research into mechanisms of DNA replication, damage and repair, we know that mutations can arise from errors in replication, such as the incorporation of a non-complementary nucleotide opposite an intact template nucleotide during DNA synthesis, or from DNA damage caused by exogenous mutagens or endogenous reactions at any time during normal growth of the cell (Figure 1). If uncorrected by the next round of DNA replication, these lesions will lead to arrested replication and cell death, or to mutations in the descendent cells (either because of incorrect template information or due to lesion bypass by error-prone DNA polymerase).

While the fraction of mutations that is non-replicative in origin remains unknown, the common assumption is that mutations are predominantly replicative (e.g., Chang et al., 1994, Li et al., 1996, Makova and Li, 2002, Tomasetti and Vogelstein, 2015). The basis for this assumption is a set of observations from disparate fields suggesting that, at least in mammals, mutations seem to track cell divisions. First, in phylogenetic studies, it has been observed repeatedly that species with longer generation times tend to have lower substitution rates, which under neutrality reflects lower mutation rates per unit time ("the generation-time effect") (e.g., Li et al., 1996, Yi et al., 2002). Second, based on comparisons of X, Y and autosomes, it has been inferred that males contribute substantially more mutations than females (e.g., Shimmin et al., 1993, Chang et al., 1994, Makova and Li, 2002). In human genetics, pedigree studies have revealed a male bias in mutation of approximately 3:1 at a paternal age of 30, and a linear increase in the number of mutations in the child with the father's age (e.g., Kong et al., 2012, Francioli et al., 2015). These observations are all qualitatively consistent with mutations arising from the process of copying DNA: all else being equal, organisms with shorter generation times should undergo more germ cell divisions per unit time and, in mammals, oocytogenesis is completed by birth whereas spermatogenesis is ongoing since puberty throughout the male lifespan, resulting in more germ cell divisions in males than females (Figure 2A).

An informative exception to the "generation time effect" seen in phylogenetic studies is transitions at CpG sites, which represent ~20% of de novo germline mutations (Kong et al., 2012), and show relatively constant substitution rates across species (Hwang and Green, 2004, Kim et al., 2006, Ségurel et al., 2014). Their more "clock-like" behavior may reflect their distinct molecular origin (Hwang and Green, 2004), as CpG transitions are believed to be due primarily to spontaneous deamination of the 5-methylcytosine (5mC) (Bird, 1980). This case demonstrates the potential importance of non-replicative sources in germline mutations, and raises the possibility that, despite the usual assumption (e.g., Taylor et al., 2006, Thomas and Hahn, 2014), not all non-CpG mutations arise from mistakes in replication.

A third argument for the preponderance of replication errors has been made recently on the basis of the observation that the lifetime risk of cancer is positively correlated with the total number of stem cell divisions across tissues (Tomasetti and Vogelstein, 2015). This finding was interpreted as indicating that more rapidly dividing tissues are accumulating more mutations through the chance occurrence of replication errors in stem cells (Tomasetti and Vogelstein, 2015). However, environmental mutagens are known to influence the incidence of a subset of cancers, indicating that mutations that have a non-replicative source can play a role in cancer risk (e.g., Irigaray et al., 2007, Parkin et al., 2011). Moreover, the sequencing of tumor samples has revealed characteristic mutation patterns ("mutational signatures") that reflect known DNA damage processes by endogenous or exogenous sources (Alexandrov et al., 2013b). These apparently conflicting observations have spurred a heated debate about the relative contributions of environment, genetics and chance to cancer risk (Song and Giovannucci, 2015, Ashford et al., 2015, Wild et al., 2015), highlighting the importance of understanding how mutations arise in somatic tissues as well as in the germline.

Because, to date, arguments for the replicative origin of mutations have been qualitative and often based on implicit assumptions, we decided to model how the source of mutations relates to their rate of accumulation over cell divisions. This model provides explicit expectations for how mutations should accumulate with sex, age and time, thus providing a single framework within which to interpret observations from evolutionary biology, human genetics and cancer genetics.

## Results

**The accumulation of mutations due to replication errors.**

The total number of mutations accumulated in one generation is the sum of mutations that arose throughout ontogeny in both parents. If mutations are introduced by replication errors, their accumulation will track rounds of DNA replication. In each developmental stage, the number of replication-driven mutations can then be expressed as the product of the number of cell divisions and per cell division mutation rate. Although a fixed per cell division mutation rate is often assumed, explicitly or implicitly (Chang et al., 1994, Drost and Lee, 1995), this need not hold, especially when the cell lineage goes through different development stages, as do germ cells of multicellular organisms. Thus, we consider a more general case, allowing for variation in per cell division mutation rate (in particular, for a higher mutation rate in early embryonic development) (Gao et al., 2011, Thomas and Hahn, 2014, Ségurel et al., 2014).

Because the process by which replication-driven mutations accumulate is not linear throughout ontogeny, the rate at which replication-driven mutations arise is highly unlikely to be strictly proportional to the generation time (Figure S1). Moreover, the average yearly mutation rate will be affected by the onset of puberty and mean age of reproduction (Figure S1, see Materials and methods equation 5). Because mammalian species differ drastically in life history traits and development and renewal processes of germ cells (Drost and Lee, 1995, Hermann et al., 2010), this simple derivation implies that the yearly mutation rate likely varies among species (even if per cell division mutation rates remain constant). As a result, in the absence of stabilizing selection on the yearly mutation rate, we should not expect neutral substitution rates to be constant across mammalian species or even along single evolutionary lineages. An important implication is that changes in life history among hominins (Robson and Wood, 2008) introduce uncertainty about dates in human evolution obtained under the assumption of a molecular clock.

Our model further reveals that, all else being equal, a longer generation time can lead to either an increase or decrease in the average yearly mutation rate, depending on parameters such as the onset of puberty, mean age of reproduction and per cell division mutation rates (Materials and methods). Therefore, the general observation that substitution rate in mammals tend to *decrease* with increasing generation times (Li et al., 1996, Yi et al., 2002, Hwang and Green, 2004) is not necessarily expected; in fact, its existence requires very specific conditions on ontogenesis to hold (see Materials and methods).

**The accumulation of non-replicative mutations with cell divisions.**
We start by considering base substitutions that result from incorrect template information caused by DNA damage prior to replication, the best understood example of which is transitions at CpG sites. This modification turns the methylated cytosine into a thymine; if uncorrected before DNA replication, an adenine will then be incorporated into the nascent strand instead of a guanine (referred to as "miscoding" lesion), which results in a mutation in one of the two daughter cells. While DNA replication and cell division are obviously two distinct events, they are tightly coordinated such that DNA is replicated exactly once before each cell division (other than in meiosis and under a few unusual conditions). In what follows, we therefore do not distinguish between the two events and assume that DNA replication is instantaneous and that the time between two consecutive rounds of replication is the same as that between two consecutive cell divisions (which is reasonable if the total lengths of G2 and M phases of the cell cycle are relatively short and constant).

We model the proportion of mutated bases at the time of cell division by considering the effects of both damage and repair (Figure 3A). For brevity, we assume that single-strand lesions and their repair occur instantaneously at constant rates, $\mu$ and $r$, respectively (Figure 3A). The proportion of base pairs that carry a lesion at time $t$ after the last cell division, $p_1(t)$, is then described by a simple differential equation:

$$\frac{dp_1}{dt} = \mu(1 - p_1) - rp_1,$$

with the initial condition $p_1(0)=0$.

Because each unrepaired single-strand lesion leads to a base pair substitution in one of the two daughter cells, the average mutation rate in one cell division (i.e., the expected fraction of base pairs that differ between a daughter cell and its mother cell) is:

$$M_{NR}(T) = \frac{\mu}{2(\mu + r)}(1 - e^{-(\mu+r)T}),$$

where $T$ is the time between two consecutive cell divisions (Figure 3B). This model can be extended readily to incorporate more features, such as other types of non-replicative mutations as well as to understand phenomena such as the strand bias in mutations associated with transcription (Green et al., 2003, Pleasance et al., 2010). Because these extensions yield qualitatively similar results (Materials and methods Section II), in what follows, we focus on the results of the simple model.

A key feature of the model is that the accumulation of mutation per cell division exhibits two different limiting behaviors, depending on the relative rates of cell division and repair. This property stems from the fact that the rate of mutation per cell division is always much smaller than the rate of cell division ($\mu T \ll 1$). It then follows that when the rate at which lesions are repaired is much slower than the rate of cell division ($rT \ll 1$), i.e., when repair is highly inefficient or the cell divides very rapidly, the number of mutations is approximately proportional to time since the last round of DNA replication (i.e., $M_{NR}(T) \approx \frac{\mu T}{2}$). The intuition is that, for a cell under this condition, there is almost no time for the repair machinery to correct lesions, so almost all lesions result in mutations. As a result, mutations accumulate at a constant rate regardless of the rates of cell division and repair (Figure 3B, red box). In other words, non-replicative mutations that are inefficiently repaired will track absolute time.

This finding partially justifies the expectation that neutral substitution rates for spontaneous mutations will not vary among species, but also highlights additional conditions for this expectation to be met. In particular, it reveals that the clock-like behavior of CpG transitions in mammals not only requires a non-replicative origin, but also implies relatively low repair efficiency in germ cells and similar damage rates across species.

Another implication is that there should be a maternal age effect on *de novo* mutation rate for CpG transitions and more generally for spontaneous mutations that are repaired inefficiently. This effect need not be strong, as oocytes may experience fewer spontaneous lesions than spermatogonial stem cells, but should be detectable with sufficient data. More generally, the detection of any maternal age effect on mutation would provide *prima facie* evidence for the existence of spontaneous mutations that are not efficiently repaired (assuming no relationship between the age at which an oocyte is ovulated and the number of cell divisions in oocytogenesis (Rowsey et al., 2014)).

In contrast, in the other limit, when the repair is highly efficient relative to the rate of cell division ($rT \gg 1$), the number of mutations approaches equilibrium levels (i.e., $M_{NR}(T) \approx \frac{\mu}{2(\mu + r)}$) by the time of division. As a result, mutations accumulate at a rate that is roughly proportional to the number of cell divisions, regardless of absolute time (Figure 3B, blue box). Here, the intuition is that when repair is highly efficient, the few lesions that have not been corrected tend to be those that arose right before the cell division, and therefore the time since the last division has little effect. Importantly, under this scenario, the rate of accrual of mutations that arise from lesions mimics what would be expected from replication errors.

In general, when repair is neither inefficient nor extremely efficient, given fixed damage and repair rates, faster dividing cells are expected to accumulate non-replicative mutations at a higher rate (per unit time) than more slowly dividing cells (Figure 3C, see Materials and methods for derivation).

## Discussion
### Implications of the model for non-replicative mutations

These results demonstrate the fundamental importance of the repair efficiency in determining the dependence of mutation rates on age, sex and time. Notably, if non-replicative mutations are efficiently repaired, their number should increase with age in males but not females (since damage in oocytes saturates). Moreover, if damage and repair rates vary across developmental stages, non-replicative mutations that are efficiently repaired may not arise in strict proportion to the number of cell divisions.

Also of note, changes to the repair efficiency (or to the division rate) could alter the sex and time dependence of mutations that arise non-replicatively; for example, decreases in repair efficiency could lead mutations that previously tracked cell division rates to depend more on absolute time. One implication is that the phylogenetic molecular clock need not run at a steady rate even for mutations due to spontaneous DNA damage.

Our results also help to interpret results from cancer studies. The total number of somatic mutations in cancer samples increases with the age of patient at diagnosis and accumulate at higher rates in fast renewing tissues (Tomasetti et al., 2013). Our modeling indicates that these observations are expected for both replicative and non-replicative mutations, so long as lesions are not inefficiently repaired in all somatic tissues. Importantly, then, the recently reported correlation between number of stem cell divisions and life time risk of cancer across tissues is consistent with both replicative and non-replicative origins of mutations, and does not provide evidence that most mutations are attributable to replication mistakes in stem cell divisions (what the authors referred to as "bad luck" in (Tomasetti and Vogelstein, 2015)). Instead, for example, they could be due to long-term exposure to a mutagen.

Similarly, neither the male bias in mutation nor the generation time effect in phylogenetics can be interpreted as providing evidence for a replication-driven mutational process, as these observations could also reflect mutations arising from residual lesions left after repair. Given these considerations, it becomes clear that, based on available data, a substantial proportion of human germline and somatic mutations—including those at non-CpG sites—may be non-replicative in origin.

**Conclusion**

In summary, we introduce a model that helps to interpret findings from mutation studies of cancer, human pedigrees and phylogenies. Although very simple, its behavior appears to be robust (see Materials and methods). By making explicit the relationship between the genesis of mutations and their accumulation over ontogeny, the model reveals the critical importance of both the source of mutations and the repair efficiency of lesions. Fitting models such as this one to growing data from diverse fields should provide a quantitative understanding of how molecular changes accumulate in tissues and over evolutionary time scales.

**Materials and methods**

**I. A model for mutations that directly arise from replication errors**

***1. The number of replication-driven mutations accumulated in one generation***

While previous characterization and modeling of germline mutations have often implicitly relied on the assumption of either a fixed mutation rate per generation or a constant mutation rate per germ cell division (e.g., Li et al., 1996, Lehtonen and Lanfear, 2014), neither of these assumptions clearly holds. In particular, variation in the replication error rate per division could arise from changes in the cellular environment throughout different development stages (e.g., the concentration of DNA polymerase and the abundance of dNTPs). In our model, we therefore consider the accumulation of replication-driven mutations—mutations that arise directly from replication errors—as a piece-wise linear process that depends on the number of cell divisions in each development stage as well as on the average mutation rate per cell division in the corresponding stage (following Ségurel et al., 2014).

Although it could be more finely graded, for simplicity, we divide the time over which replication-driven mutations accrue in germ cells into four stages: the stage from fertilization to the settlement of primordial germ cells (PCGs) in the developing gonads (which almost coincides with sexual differentiation); from PGC settlement to birth; from birth to the onset of puberty; and from puberty to reproduction. Let $d_i^s$ and $\mu_i^s$ be the numbers of cell divisions and replication error rate in the $i^{th}$ stage ($i$ =1, 2, 3, 4) in sex $s$ ($s \in \{f, m\}$). Because sexual differentiation does not take place until the second stage, $d_1^f = d_1^m$ and $\mu_1^f = \mu_1^m$, which can be replaced with simply $d_1$ and $\mu_1$. Previous studies in *Drosophila melanogaster* suggest that the first division of a zygote has an extraordinarily high mutation rate (Gao et al., 2011, Gao et al., 2014). Although the first division in Drosophila is quite distinct from that in mammals, it seems possible that it would be more mutagenic in mammals as well, and we therefore consider the first division separately as stage 0, of which the mutation rate is $\mu_0$ for both sexes, and re-define stage 1 as from the second post-zygotic division to sex differentiation. The total number of replication-driven autosomal mutations from one parent to the offspring is then:

$$M_R^s = (\mu_0 + \mu_1 d_1 + \mu_2^s d_2^s + \mu_3^s d_3^s + \mu_4^s d_4^s)H, \ s \in \{f, m\}$$

where $H$ is the total number of base pairs in a haploid set of autosomes.

In mammals (and birds), all mitotic divisions of female germ cells are completed by birth of the future mother, so $d_3^f \approx 0$ and $d_4^f = 0$, and the total number of replication-driven mutations inherited from mother is (Figure 2B red line):

$$M_R^f = (\mu_0 + \mu_1 d_1 + \mu_2^f d_2^f)H. \quad (1)$$

We note that although experimental evidence in mice suggests that late-ovulated oocytes tend to undergo more cell divisions than early-ovulated oocytes (Reizel et al., 2012), there is evidence against the production-line hypothesis in humans (Rowsey et al., 2014). Since we focus on humans, throughout we assume that there is no relationship between the age of reproduction of the mother and the number of cell divisions that led to the oocyte.

In males, in contrast, germ cells undergo divisions in all stages outlined above; furthermore, the number of germ cell divisions after puberty ($d_4^m$) is not a fixed number, because sperm are continuously produced through asymmetric division of spermatogonial stem cells (SSCs) at an approximately constant rate since puberty. If we assume that males and females have the same ages of puberty and reproduction (denoted by $P$ and $G$ respectively), and that SSCs undergo $c^m$ cell divisions each year, the total number of cell divisions from puberty to completion of spermatogenesis is roughly:

$$d_4^m = c^m(G - P - t_{sg}) + d_{sg},$$

where $t_{sg}$ and $d_{sg}$ are the time (in years) and the number of cell divisions needed to complete spermatogenesis from SSC. The two divisions in meiosis are counted as one here, because only one round of DNA replication takes place. Therefore, the total number of paternal mutations is a function of reproductive age $G$ (Figure 2B blue line):

$$M_R^m = [\mu_0 + \mu_1 d_1 + \mu_2^m d_2^m + \mu_3^m d_3^m + \mu_4^m(c^m(G - P - t_{sg}) + d_{sg})]H. \quad (2)$$

Summing equations (1) and (2), the total number of autosomal replication-driven mutations inherited by a diploid offspring from both parents is (Figure 2B purple line):

$$M_R = M_R^f + M_R^m$$
$$= [2\mu_0 + 2\mu_1 d_1 + \mu_2^f d_2^f + \mu_2^m d_2^m + \mu_3^m d_3^m + \mu_4^m(c^m(G - P - t_{sg}) + d_{sg})]H. \quad (3)$$

### 2. The per generation mutation rate and the average yearly mutation rate

It follows from equation (3) that the mutation rate per basepair per generation is an increasing function of $G$ (Figure S1):

$$m_{R,g} = \frac{M_R}{2H} = [2\mu_0 + 2\mu_1 d_1 + \mu_2^f d_2^f + \mu_2^m d_2^m + \mu_3^m d_3^m + \mu_4^m d_{sg} + \mu_4^m c^m(G - P - t_{sg})]/2.$$

From the equation above, we can further obtain the average yearly mutation rate (i.e., the substitution rate if all mutations are neutral):

$$m_{R,y} = \frac{m_{R,g}}{G} = \frac{2\mu_0 + 2\mu_1 d_1 + \mu_2^f d_2^f + \mu_2^m d_2^m + \mu_3^m d_3^m + \mu_4^m d_{sg} + \mu_4^m c^m (G - P - t_{sg})}{2G}. \quad (4)$$

By dividing equation (2) by equation (1), we obtain the ratio of male to female replication-driven mutations:

$$\alpha_R = \frac{M^m}{M^f} = \frac{\mu_0 + \mu_1 d_1 + \mu_2^m d_2^m + \mu_3^m d_3^m + \mu_4^m d_{sg}}{\mu_0 + \mu_1 d_1 + \mu_2^f d_2^f} + \frac{\mu_4^m c^m}{\mu_0 + \mu_1 d_1 + \mu_2^f d_2^f} \cdot (G - P - t_{sg}),$$

which suggests that, keeping other parameters unchanged, increases in generation time $G$ will lead to a stronger male bias in mutation, as expected intuitively (Figure 2C).

### *3. The effect of generation time on the yearly mutation rate*

Increases in $G$ lead to increases in both the numerator and the denominator, so it is unclear whether the yearly mutation rate will increase or decrease with $G$ from this equation alone. To explore the effect of generation time on the average yearly mutation rate, it is useful to reorganize equation (4) as:

$$m_{R,y} = \frac{\mu_4^m c^m}{2} + \frac{2\mu_0 + 2\mu_1 d_1 + \mu_2^f d_2^f + \mu_2^m d_2^m + \mu_3^m d_3^m + \mu_4^m d_{sg} - \mu_4^m c^m (P + t_{sg})}{2G}$$

$$= \frac{\mu_4^m c^m}{2} + \frac{A^*}{2G}, \quad (5)$$

where $A^* = 2\mu_0 + 2\mu_1 d_1 + \mu_2^f d_2^f + \mu_2^m d_2^m + \mu_3^m d_3^m - \mu_4^m (c^m P + c^m t_{sg} - d_{sg})$.

Equation (5) suggests that if and only if $A^*=0$ will the yearly mutation rate be independent of $G$. Otherwise, $m_{R,y}$ will either monotonically increase or decrease with $G$, depending on the sign of $A^*$. Changes in the timing of puberty ($P$), in the number of cell divisions ($d_i^s$) and in the replication error rate per cell division in each stage ($\mu_i^s$) will also influence the value of $m_{R,y}$ and its dependence on $G$.

The relationship between $m_{R,y}$ and $G$ can also be directly read off the curve given by equation (3) (Figure S1). The total number of mutations increases linearly with $G$ after puberty, but this linear relationship does not apply to the period before puberty. If (and only if) the extended line passes through the origin will the total number of mutations be exactly proportional to the generation time, and the average yearly mutation rate unaffected by $G$. If the intercept of the extrapolated line at age zero is positive, $m_{R,y}$ *decreases* with $G$, the direction of the observed "generation time effect" in primates. Conversely, if the intercept is negative, $m_{R,y}$ *increases* with $G$. In fact, the intercept obtained by extrapolation is exactly $A*H$, so the interpretation from the curve is equivalent to that suggested by equation (5).

Although estimates of other parameters exist, little is known about the replication error rate per cell division in germ cells, so it is unclear whether $A^*$ is positive or negative. However, it seems highly coincidental that an expression that involves multiple variables would happen to equal to zero. Therefore, we argue that, in absence of strong selection constraints on the yearly mutation rate, there is almost certainly an effect of generation time on yearly mutation rate in humans, although the effect could be weak (Ségurel et al., 2014).

## 4. Properties of germline mutations inferred from human trio studies suggest the yearly mutation rate should decrease with generation time

One way to test whether there exists a generation time effect in humans is to extend the fitted line of empirical *de novo* mutation data in order to estimate the intercept at age zero. When we combine data from three available whole-genome datasets (Campbell et al., 2012, Michaelson et al., 2012, Jiang et al., 2013, Ségurel et al., 2014) in order to consider the effect of paternal age on mutation rates, the intercept of the linear regression line is significantly positive. This suggests that a generation time effect might be operating in humans (at least in the populations under study).

The positive intercept also has interesting implications for the per cell division mutation rates in different development stages of germ cells in humans. As shown above, there is a positive intercept if and only if:

$$2\mu_0 + 2\mu_1 d_1 + \mu_2^f d_2^f + \mu_2^m d_2^m + \mu_3^m d_3^m > \mu_4^m (c^m P + c^m t_{sg} - d_{sg}).$$

It is estimated that, in humans, $d_1$=15, $d_2^f$=15, $d_2^m$=21, $d_3^m$=0, $P$=13, $t_{sg}$=0.2, $d_{sg}$=4 and $c^m$=23 (Drost and Lee, 1995, Nielsen et al., 1986). By plugging in these estimates, we obtain:

$$2\mu_0 + 30\mu_1 + 15\mu_2^f + 21\mu_2^m > (23 \times 13.2 - 4)\mu_4^m = 303.6\mu_4^m. \quad (6)$$

Because the coefficient in front of $\mu_4^m$ is so large, inequality (9) cannot hold if $\mu_0$, $\mu_1$, $\mu_2^f$ and $\mu_2^m$ are all smaller than or equal to $\mu_4^m$. In other words, at least one of the four pre-puberty per cell division mutation rates has to be much larger than the post-puberty rate in males.

One possibility is that the first division in the zygote has an extremely high mutation rate, while all other cell divisions share similar mutation rates, as appears to be the case in a completely different developmental context, in *Drosophila melanogaster* (Gao et al., 2014, Gao et al., 2011). If we assume that $\mu_1 = \mu_2^f = \mu_2^m = \mu_4^m$, then inequality (6) reduces to:

$$\mu_0 > 117\mu_1,$$

which is surprisingly high, but on par with what is seen in *D. melanogaster*, where the first post-zygotic division is estimated to be ~800 fold more mutagenic than other development stages of the embryogenesis (Gao et al., 2011).

In addition to an extraordinarily high mutation rate for the first cell division, other scenarios (for example, $\mu_0 = \mu_1 = \mu_2^f = \mu_2^m > 4.46\mu_4^m$) could also explain a positive intercept at age zero. More data from humans is needed to distinguish between these scenarios. Also needed are better estimates of numbers of germ cell divisions in each development stage in humans, as current knowledge about human spermatogenesis largely comes from one study in the early 1960s, which determined SSC division cycle by radioautograph of testicular biopsies after injection of tritiated thymidine (Heller and Clermont, 1963).

### *5. Properties of germline mutations inferred from a chimpanzee pedigree suggest that the yearly mutation may increase with generation time*

A recent pedigree study in Western chimpanzees provides the opportunity to evaluate the presence and magnitude of a generation effect on de novo mutation rate in a non-human great ape species. Intriguingly, the intercept of mutation accumulation curve at age zero appears to be negative. While it is not significantly less than 0, if this finding were to hold up with more data, it would imply that the substitution rate should stay relatively constant or even *increase* with generation time, i.e., that chimpanzees would show the opposite of the generation time effect. It is thought that chimpanzees reach puberty at earlier age ($P$=8.75 since conception) (Behringer et al., 2014) and have shorter duration of spermatogenesis and spermatogenic cycles compared to humans ($t_{sg}$=0.17, $c^m = 26$) (Smithwick et al., 1996), but data on the number of cell divisions in each development stage are much more scarce. If we assume that the development process is conserved between humans and chimpanzees until birth and use the estimated numbers of cell divisions in humans as proxies for those of chimpanzees, the non-significant negative intercept at age zero indicates that:

$$2\mu_0 + 30\mu_1 + 15\mu_2^f + 21\mu_2^m \leq (26 \times 8.58 - 4)\mu_4^m = 219\mu_4^m. \qquad (7)$$

Taking these point estimates at face value, the comparison between the results of studies in the two species leads to two intriguing findings. First, comparing (6) and (7) suggests that the mutation rates per cell division cannot all be the same for the two species under the assumption of conserved germ cell development process, indicating that either the mutation rates or the numbers of cell divisions during various stages have evolved between these two closely related species. Second, despite distinct onsets of puberty and the different relationships between the average generation time and the number of *de novo* mutations, the average yearly mutation rates are remarkably similar for humans and Western chimpanzees. However, the sample size for the chimpanzee study is very small, and the captive animals' reproductive ages differ substantially from wild animals, so further data are required to assess if these conclusions are solid. What is underscored by this comparison, however, is that a decrease in the mutation rate with increasing generation time (i.e., the "generation time effect") is by no means a given, and in fact requires quite specific conditions to hold.

**II. A model for mutations that arise from DNA lesions**
***1. The basic model for non-replicative mutations***
As many as 50,000 DNA damaging events are thought to occur in each cell every day as a result of normal cellular metabolism, and more DNA lesions may be generated by exogenous agents (Salk et al., 2010). Typical DNA damage includes depurination and deamination due to DNA hydrolysis; alkylation and oxidation of bases induced by chemicals such as ethylmethane sulfonate (EMS) or reactive oxygen species (ROS); pyrimidine dimers caused by ultraviolet (UV) radiation; and single- or double-stranded breaks produced by gamma and X-rays. Most of those single-stranded lesions cannot pair properly with any base (termed "noncoding bases") and thus will block DNA replication if unrepaired (Figure 1 in the main text). However, a few alterations to nucleotides can pair with bases different from the original Watson-Crick partners; such lesions (termed "miscoding bases"), if left unrepaired before replication, will lead to irreversible replacement of a base pair after cell division (Figure 1). The spontaneous deamination of 5mC at CpG sites is a typical example of such lesions. We start by focusing on this scenario and discuss more complex mutagenesis mechanisms later.

Since mutation results from a process that involves damage, repair and replication, we model the proportion of mutated bases at the time of cell division by considering the effects of both damage and repair. For simplicity, we assume that the amounts of mutagens (endogenous or exogenous, chemical or physical) and repair enzymes are constant throughout the cell cycle, so single-strand damage occurs at a constant instantaneous rate $\mu$; we further assume that the repair machinery recognizes lesions at a constant instantaneous rate $r$ (Figure 3A). Therefore, each DNA base pair in the genome can be in one of two states at any given time: with or without a lesion. We term these two states $S_0$ and $S_1$ and use $p_0(t)$ and $p_1(t)$ to denote the proportions of base pairs in each state at time $t$ since the last cell division. Therefore, the initial condition is $p_0(0)=1$ and $p_1(0)=0$, and the dynamics of the two proportions can be described with the differential equations:

$$\frac{dp_0}{dt} = -\mu p_0 + r p_1,$$

$$\frac{dp_1}{dt} = \mu p_0 - r p_1.$$

In this case, the system is easily reducible to a single equation as $p_0(t) + p_1(t) = 1$ (cf. main text), and we use a system only to underline the relationship between the simplified model and the generalizations in Section II.4.

If a base pair is in state $S_0$ at the time of cell division, there is no sequence change in the daughter cells; in contrast, if a base pair is in state $S_1$, the lesion leads to a base pair replacement in one of the two daughter cells. Therefore, the effective mutation rate per base pair for this cell division (i.e., the average fraction of DNA base pairs that differ between the mother cell and one of its daughter cells) is $M_{NR}(T) = ½ p_1(T)$. We assume that $\mu << 1/T$ for any biologically reasonable value of $T$, so even in the absence of DNA repair, the absolute mutation rate per division (½$\mu T$) is very small. In addition, we assume an infinite sites model, in which each genomic site can be mutated at most once. Thus, the mutation rate over many cell divisions is the sum of the mutation rates over every division.

## 2. Mutation rate per cell division

The solution to the differential equation is:

$$p_1(t) = \frac{\mu}{\mu + r}(1 - e^{-(\mu+r)t}), \text{ and } p_0(T) = 1 - p_1(T) \text{ for any } T>0.$$

Therefore, the mutation rate per cell division is a function of the time between two consecutive cell divisions $T$:

$$M_{NR}(T) = \frac{\mu}{2(\mu + r)} (1 - e^{-(\mu+r)T}), \quad (8)$$

which can be re-written as $M_{NR}(T) = \frac{1}{2(1+R)}(1 - e^{-(1+R)\mu T})$ for greater clarity,

where $R = \frac{r}{\mu}$ measures the relative rate of repair compared to the rate of damage.

When repair is highly inefficient or the time between two cell divisions is very short (i.e., $rT<<1$), we have:

$$M_{NR}(T) \approx \frac{\mu T}{2}. \quad (9)$$

In other words, when $rT<<1$, the number of mutations is approximately proportional to absolute time and independent of the cell division or repair rate. We term this part of the curve the "linear phase". An intuitive way to understand this finding is that, when repair is inefficient compared to the cell division rate, there is almost no time for the repair machinery to correct DNA lesions, so almost all single-strand lesions result in mutations in one of the two daughter cells after cell division, and mutations accrue approximately linearly with time. This result suggests that lesions that have the same damage rates but are recognized by different repair mechanisms may show distinct time dependencies, as well as differences in absolute mutation rates.

On the other hand, when repair is highly efficient relative to the rate of cell division (i.e., $rT>>1$), the mutation rate per cell division approaches a limit:

$$M_{NR}(T) \approx \frac{1}{2(1+R)}. \quad (10)$$

We term this part of the curve the "asymptotic phase" (for which $rT>>1$, which implies $r>>\mu$). Interestingly, the limiting value only depends on the ratio of $r$ to $\mu$, regardless of their absolute values. The intuition is that, when the repair takes place sufficiently rapidly, the cell approaches an equilibrium between damage and repair before each division; hence, the number of unrepaired lesions before each cell division is roughly the same and depends only on the relative magnitudes of the damage and repair rates. In the asymptotic phase, non-replicative mutations accrue with cell division rate, which mimics what would be expected for mutations that result from replication mistakes. We note that the existence of such an equilibrium comes from the assumption of no error in repair); however, the main conclusions hold when extending the model to incorporate errors (see Section II.4).

In between these two limiting behaviors, the mutation rate per cell division will increase with the time between two consecutive cell divisions ($T$), but the rate of this increase decreases (Figure 3B). In other words, $M_{NR}(T)$ is a concave function of $T$, which is also evident from its first and second derivatives with regard to $T$:

$$M_{NR}'(T) = \frac{\mu}{2} e^{-(1+R)\mu T} > 0,$$

$$M_{NR}''(T) = \frac{-(1+R)\mu^2}{2} e^{-(1+R)\mu T} < 0.$$

As a result of the concavity, the mutation accumulation curve will progressively deviate from the linear expectation with no repair.

### 3. Mutation rate per unit time

To understand how the total number of mutations accumulated in a given period of time (e.g., by the same age) depends on the rate of cell division (e.g., in different tissues), we derive the mutation rate per unit time as the product of mutation rate per cell division and the rate of cell division ($c=1/T>0$):

$$m(c) = cM_{NR}(\frac{1}{c}) = \frac{c}{2(1+R)}(1 - e^{-\frac{(1+R)\mu}{c}}). \qquad \text{(Figure 3C)}$$

The mutation rate $m(c)$ has two limiting behaviors when $c$ approaches infinity and zero, which have the same intuitive explanations as equations (9) and (10), respectively. Moreover, it can be shown that $m(c)$ is a concave increasing function of $c$. In other words, in a given period of time, faster dividing cells accumulate more non-replicative mutations than slowly dividing cells, but the increase in the number of mutations is smaller than the increase in the cell division rate.

### 4. More complex scenarios of mutagenesis

The analyses above rely on several simplifying assumptions about the DNA damage, repair and replication processes. Here, we briefly discuss how the basic model can be extended to model more complex scenarios of mutagenesis and to understand more detailed patterns of mutations, such as the strand bias in mutation rates associated with transcription (Green et al., 2003, Pleasance et al., 2010). Importantly, the analyses show that key qualitative behaviors of the basic model presented above and in the main text are robust to these extensions.

(1) Other miscoding lesions

Although the basic model was presented as describing deamination of 5-mC at CpG, it also applies to other types of miscoding lesions. If a miscoding lesion is not repaired correctly before DNA replication, it leads to the incorporation of an incorrect base in the newly synthesized strand. This results in a "semi-substitution" of the base pair after cell division. Although afterwards, the modification itself may still be recognizable by the DNA repair pathway, it is unlikely to be repaired correctly, because the original template is lost and substituted by another regular DNA base. Therefore, unrepaired miscoding lesions will almost always lead to mutations, just as in the case of deamination of 5-methylcytosine. Hence, our model can be applied to other miscoding lesions, such as O-6-ethylguanine and 8-oxoguanine that result from alkylation and oxidation of guanine, respectively (Friedberg, 2006), and the accrual of mutations will have the same behavior as described above (Figure 3B,C).

(2) Noncoding lesions

The model also describes the accumulation of mutations that result from noncoding lesions, i.e., abasic sites or modified DNA bases that cannot pair with any regular nucleotides. When the DNA replication machinery encounters such a lesion, it will either stall and then trigger cell death or bypass the lesion by cleaving the "irregular" base and incorporating a random base instead. In the latter case, the incorporated base is likely to differ from the original one and will result in a mutation in one of the two daughter cells. The probability that the randomly incorporated base happens to be the original one will depend on the concentrations of the four dNTPs in the cell, so the per cell division mutation rate is:

$$M_{NR}(T) = \tfrac{1}{2} K p_1(T), \quad\quad\quad\quad (11)$$

where $K$ ($0<K<1$) is a scale factor depending on the specific type of noncoding lesion. Therefore, $M_{NR}(T)$ has similar limiting behaviors as equation 8 (i.e., the linear phase and asymptotic phase).

There is one complication under this scenario though: because noncoding lesions can trigger cell death, the surviving cell lineages will carry fewer mutations on average than the expectation in equation (11). As a result, our model for miscoding lesions will tend to over-estimate the mutation rate for noncoding lesions; however, there should still exist two limiting phases. The linear phase is approximately the same as described in (11), because cell death is rarely triggered when a cell carries very few lesions (Otterlei et al., 2000). The asymptotic phase will have a lower limit value though, as cells with a greater number of noncoding lesions will be more likely to die.

(3) Different repair rates for the transcribed and non-transcribed strands

Some types of mutations display substantial asymmetry between the transcribed and non-transcribed (coding) strands, as observed in both phylogenetic and cancer studies. In sequence divergence between humans and chimpanzees, A→G transitions on the transcribed strand are 58% more common than the complementary T→C transitions (Green et al., 2003). This phenomenon can be explained by transcription-coupled repair (TCR), a sub-pathway of nucleotide excision repair (NER) that corrects errors on the transcribed strand of actively expressed genes (Hanawalt and Spivak, 2008), or by transcription associated mutagenesis (TAM) (Jinks-Robertson and Bhagwat, 2014). In lung adenocarcinoma, G→T mutations, a signature associated with tobacco carcinogens, are found to be less prevalent on the transcribed strand; in turn, in malignant melanoma, C→T mutations, which likely result from exposure to UV radiation, also occur less often on the transcribed strand (Alexandrov et al., 2013a, Pleasance et al., 2010). These observations are also consistent with TCR.

Our model can be extended to incorporate TCR or TAM by modeling the accumulation of mutations on the two strands separately, where each strand has its own damage and repair rates. Notably, because damage rates are extremely small, we can assume that each strand experiences lesions independently, so the total number of mutations is the sum of unrepaired lesions on both strands by cell division. Assuming independence between the two strands, the results about the relationship between the efficiency of repair and the accumulation of mutations apply to both strands separately.

The model can then be used to infer parameters of interest, such as the relative difference in repair rates between the transcribed and non-transcribed strands. As an illustration, we assume that all or most mutations are non-replicative and that the strand bias is completely attributable to TCR. Assuming that the two strands have the same damage rate but different relative repair rates $R_1$ and $R_2$ ($R_1 > R_2 > 0$), the per cell division mutation rates for the two strands are:

$$M_{NR,1}(T) = \frac{1}{2(1+R_1)}(1-e^{-(1+R_1)\mu T}), \text{ and } M_{NR,2}(T) = \frac{1}{2(1+R_2)}(1-e^{-(1+R_2)\mu T}).$$

If the fraction of mutations that accrue on the transcribed strand is $x$ ($x < ½$), then:

$$\frac{x}{1-x} = \frac{M_1(T)}{M_2(T)} = \frac{1+R_2}{1+R_1} \cdot \frac{1-e^{-(1+R_1)\mu T}}{1-e^{-(1+R_2)\mu T}} > \frac{1+R_2}{1+R_1} > \frac{R_2}{R_1},$$

which can be reorganized as:

$$\frac{R_1}{R_2} > \frac{x}{1-x}. \qquad (12)$$

The right-hand term in equation (12) provides a good approximation for the ratio of $R_1$ to $R_2$ when the repair rates are much greater than the damage rate ($R_1 > R_2 >> 1$) and repair is efficient on both strands (i.e., $(1+R_2)\mu T$ is moderately large). Thus, in principle, the model could be used to make inferences about the relative rates of TCR.

(4) Repair with occasional errors

To take into account the repair machinery's occasional failure to resolve mismatches correctly, we can include a repair error rate, $\varepsilon$ ($\varepsilon > 0$) in the model and assume that, when the repair is incorrect, it alters the strand without a lesion and leads to a base pair replacement. In this model, each DNA base pair can therefore be in one of three states: the original state, a state in which one strand has a lesion and one in which both strands changed (i.e., a base pair replacement). We denote the third state by $S_2$, and the proportion of base pairs in this state at time $t$ since last cell division by $p_2(t)$ (Figure S2 A). Therefore, the initial conditions are $p_0(0)=1$ and $p_1(0)=p_2(0)=0$, and the dynamics of the three proportions follow the differential equations:

$$\frac{dp_0}{dt} = -\mu p_0 + (1-\varepsilon) r p_1,$$

$$\frac{dp_1}{dt} = \mu p_0 - r p_1,$$

$$\frac{dp_2}{dt} = \varepsilon r p_1.$$

where $p_0(t) + p_1(t) + p_2(t) = 1$. If a base pair is in state $S_2$ at the time of cell division, it leads to a base pair substitution in each of the two daughter cells. Therefore, the effective mutation rate per division is $M_{NR}(T) = ½ \, p_1(T) + p_2(T)$.

The solution to this differential equation system is:

$$\vec{p}(t) = \vec{\eta}_1 e^{\lambda_1 t} + \vec{\eta}_2 e^{\lambda_2 t}, \qquad (13)$$

where $\lambda_1$ and $\lambda_2$ are the two eigenvalues of the matrix $\begin{bmatrix} -\mu & (1-\varepsilon)r \\ \mu & -r \end{bmatrix}$, and $\vec{\eta}_1$ and $\vec{\eta}_2$ are the corresponding vectors that satisfy $\vec{\eta}_1 + \vec{\eta}_2 = \begin{pmatrix} 1 \\ 0 \end{pmatrix}$, because of the initial condition.

It can be shown that:

$$\lambda_1 + \lambda_2 = -(\mu + r);$$

$$\lambda_1 \lambda_2 = \varepsilon \mu r.$$

Therefore, $\lambda_1$ and $\lambda_2$ are both negative numbers, and under the assumption that $\varepsilon$ is small, one of the two eigenvalues is very close to zero, and the other close to $-(\mu+r)$. Without loss of generality, let us assume that $\lambda_1$ is the one close to zero. Because of the similarity in form between equations (13) and (8), we predict that, when the error rate of repair is small, the results will be qualitatively similar to what we obtained from the simpler model. Indeed, we find that, for various small values of $\varepsilon$ ($\varepsilon$ =0.0001, 0.001 or 0.01), the behaviors of $p_0(t)$, $p_1(t)$ and $M_{NR}(T)$ at short time scales are very similar to the case with perfect repair (Figure S2 B). However, when $\varepsilon>0$, errors during repair will lead to accumulation of sites in state $S_2$, and $M_{NR}(T)$ will eventually approach 1 when $T$ is sufficiently large.

The mutation accumulation curve can be roughly divided into four phases (Figure S2 C):

**Phase 1 (the linear phase)**: When $|\lambda_1|t \ll |\lambda_2|t \ll 1$ (or approximately $(\mu+r)t\ll1$), $p_1(t)$ increases approximately linearly and is the main contributor to $M_{NR}(t) = \frac{1}{2} p_1(t) + p_2(t)$, because there are too few base pairs in state $S_2$ to contribute to the mutation rate. As long as $\varepsilon$ is small, the specific value of $\varepsilon$ does not significantly affect the length of phase 1 and the behaviors of $p_0(t)$, $p_1(t)$ and $M_{NR}(T)$ in this phase. Therefore, equation (9) is a good approximation: $M_{NR}(T)$ increases linearly with $T$, and the mutation rate per unit of time is independent of cell division rate.

**Phase 2 (the asymptotic phase)**: When $\varepsilon r t \ll 1 \ll |\lambda_2|t$ (or approximately $1/(\mu+r)\ll t\ll 1/\varepsilon r$), $p_1(t)$ reaches an asymptote, and the contribution of $p_2(t)$ to the mutation rate is still very small compared to that of $p_1(t)$. Therefore, equation (10) is a good approximation for this phase, which means that $M_{NR}(T)$ is relatively constant irrespective of $T$, and the mutation rate per unit of time is proportional to the cell division rate. The existence of $\varepsilon$ does not impact the behaviors of $p_0(t)$, $p_1(t)$ and $M_{NR}(T)$ in this phase, but the value of $\varepsilon$ determines the length of this phase: smaller $\varepsilon$ leads to a longer asymptotic phase due to later onset of the next phase.

**Phase 3 (the takeoff phase)**: when $|\lambda_1|t \ll 1 \ll \varepsilon r t$, $p_1(t)$ is still relatively constant with $t$, but the linear increase in $p_2(t)$ makes lesions in state $S_2$ the main source of mutations in this phase. Therefore, $M_{NR}(t)$ takes off from the temporary asymptote in phase 2 and increases linearly with $t$ again. Because $|\lambda_1| = \frac{\varepsilon \mu r}{|\lambda_2|} \approx \frac{\varepsilon \mu r}{\mu + r}$, the two boundaries of phase 3 are both proportional to $\varepsilon$, so the logarithm of its length is relatively constant. In this phase, $M_{NR}(T)$ increases linearly with $T$ but is not proportional to $T$, so the mutation rate per unit of time depends on the cell division rate but not in strict proportion.

**Phase 4 (the final phase)**: when $1 \ll |\lambda_1|t \ll |\lambda_2|t$, both $p_0(t)$ and $p_1(t)$ quickly drop to 0, and most base pairs in the genome are substituted (in state $S_2$). Therefore, the mutation rate is approximately 1 in this phase. Although this regime is mathematically possible, it is almost certainly biologically irrelevant, given that the mutation rate per cell division is very low.

In summary, other than in the linear phase where repair is highly inefficient (i.e., $(\mu+r)t \ll 1$), the rate of accrual of mutations per unit time will depend on the cell division rate. Thus, our key points highlighted the main text still hold even when the error rate is not zero but reasonably small.

(5) Varying damage rate and/or repair rate

Although we assume that the damage and repair rates are constant throughout time, in reality, they may vary considerably due to changes in chromatin state, concentrations of chemicals, and changes in expression levels of repair complex components with age (e.g., the CpG methylation level (Horvath, 2013)) or even within a cell cycle. To take into account changes in damage and repair rates with age, we can calculate the expected mutation rate by treating the accumulation of mutations as a piece-wise linear process and summing across all ages to obtain the mutation rate at any given age. To deal with varying parameters within a cell cycle, we can rewrite the differential equations by treating the damage and repair rates as functions of time since last cell division instead of constants:

$$\frac{dp_0}{dt} = -\mu(t)p_0 + r(t)p_1,$$

$$\frac{dp_1}{dt} = \mu(t)p_0 - r(t)p_1,$$

where $\mu(t)$ and $r(t)$ are known functions.

The differential equation can be solved as:

$$p_1(t) = \frac{\int e^{U(t)+R(t)} \mu(t)dt - \frac{\mu(0)}{\mu(0)+r(0)}}{e^{U(t)+R(t)}},$$

where $U(t) = \int_0^t \mu(t)dt$, and $R(t) = \int_0^t r(t)dt$. The mutation rate per cell division is $M_{NR}(T)$ = ½ $p_1(T)$.

Whether there exists a linear phase (or an asymptotic phase) depends on the specific forms of $\mu$(t) and $r$(t). Nonetheless, $M_{NR}(T)/T$ is unlikely to be approximately constant, because its first derivative is a complex function involving $\mu$(t) and $r$(t) and is unlikely to be zero for any *t*. Therefore, the cell division rate will affect the mutation rate per unit time.

If $\mu$(t) and $r$(t) are well characterized, the mutation rate can be solved numerically using Equation (20). Therefore, this extended model can be used to describe how mutations with varying damage and repair rates accumulate with age, sex and time.

**Acknowledgements**

We thank Phil Green for comments on a previous paper that motivated us to pursue this work; Eduardo Amorim, Guy Amster, Chen Chen, Priya Moorjani and Laure Ségurel for helpful discussions; and Ludmil Alexandrov, Michael Stratton and Shamil Sunyaev for sharing unpublished data and valuable suggestions. Z.G. was supported in part by the William Rainey Harper Fellowship from the University of Chicago.

**Figure 1. An overview of the mutagenesis process, which involves DNA damage, repair and replication.**

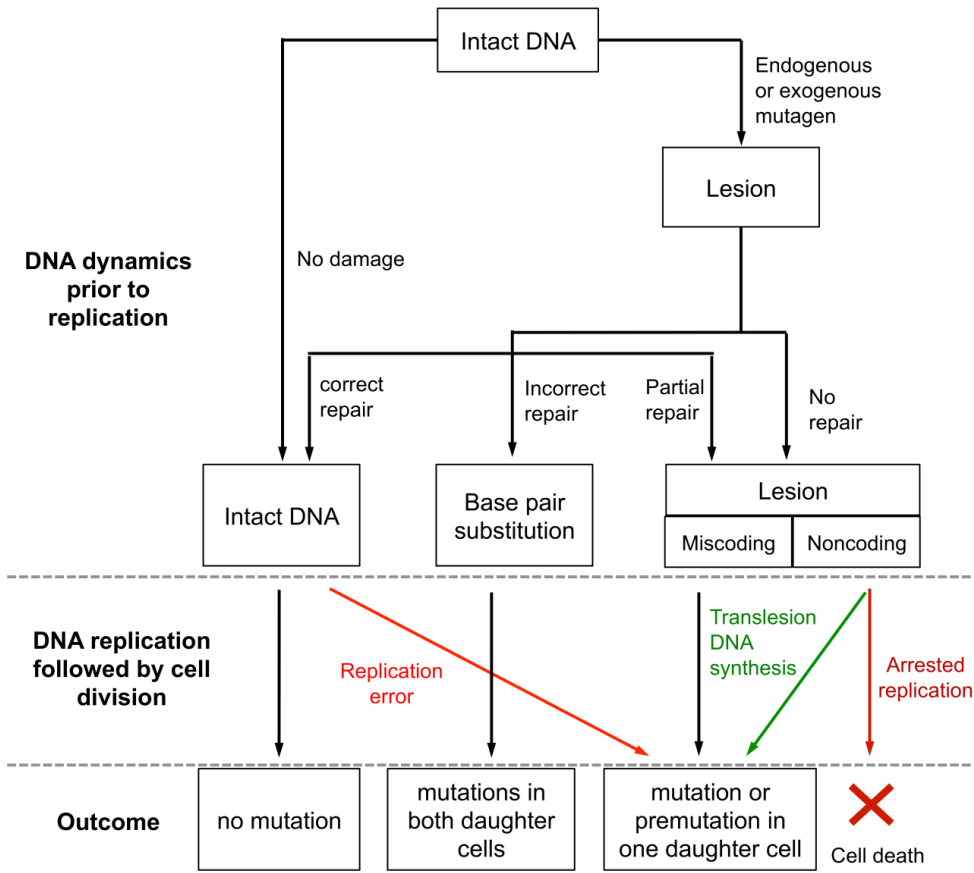

**Sidebar: Explanation of the terms used in Figure 1.**

| Term | Explanation |
|---|---|
| Lesion | Chemically altered base |
| Noncoding lesion | Lesion that cannot pair properly with any regular DNA bases |
| Miscoding lesion | Lesion that pairs with regular DNA bases that differ from the original one |
| Correct repair | Repair that completely reverses the lesion to the original state |
| Incorrect repair | Repair that recognizes the mismatch caused by lesion but alters the undamaged base by mistake |
| Partial repair | Incomplete repair that leads to abasic sites or other base alterations |
| Replication error | Mis-incorporation of nucleotide in the newly synthesized strand despite intact template |
| Translesion DNA synthesis | Damage tolerance mechanism that allows the DNA replication to bypass lesions and is often mutagenic |
| Point mutation | Base pair substitution |
| Premutation | A base pair at which a lesion is present on one strand and the base on the other strand is substituted, which results from replication according to incorrect template information |

**Figure 2. The accumulation of replication-driven mutations with sex and age.**

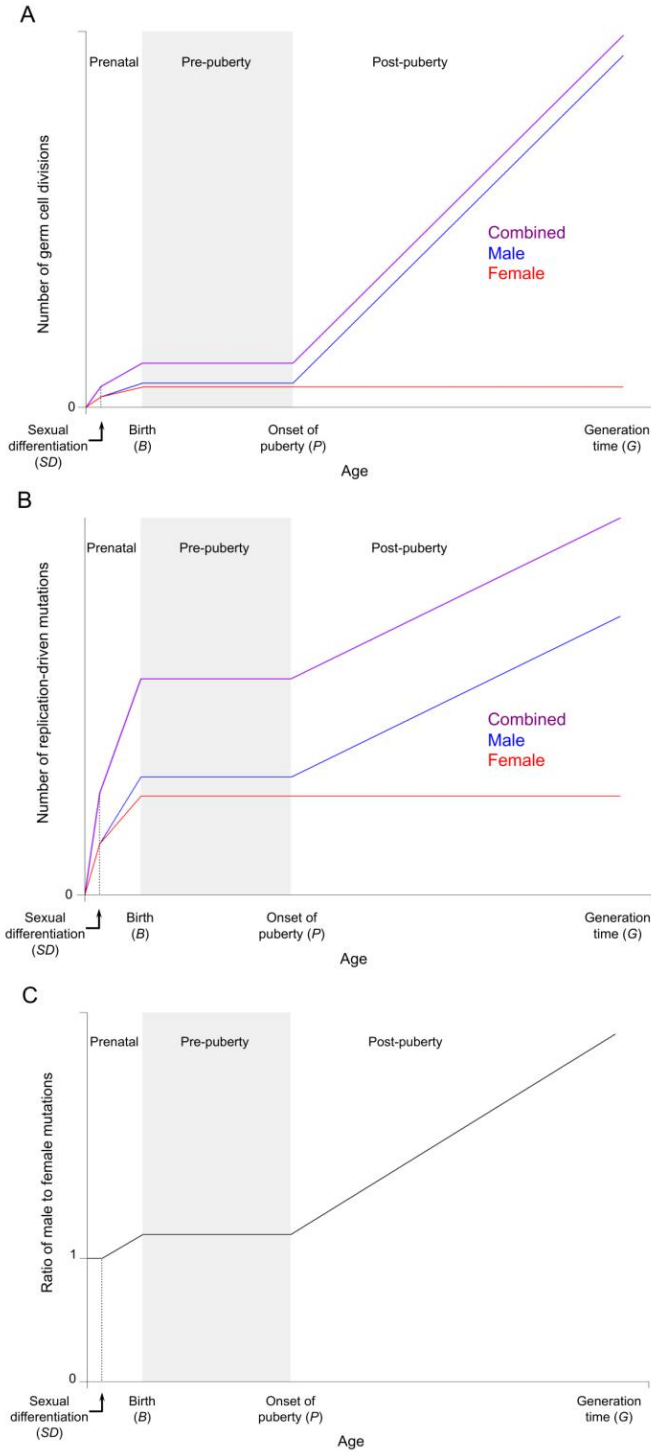

A) A cartoon of the increase in the number of germ cell divisions with age, in humans. For legibility, the plot is not exactly to scale and the four cell divisions needed to complete spermatogenesis are not shown. The origin is the time of fertilization and SD, B, P and G the times of sexual differentiation, birth, onset of puberty and reproduction (i.e., generation time), respectively.
B) The increase in the number of mutations due to replication errors with sex and age.
C) The ratio of mutations that occurred in the male versus the female germline (the "male bias") with age.

**Figure 3. The basic model for non-replicative mutations.**

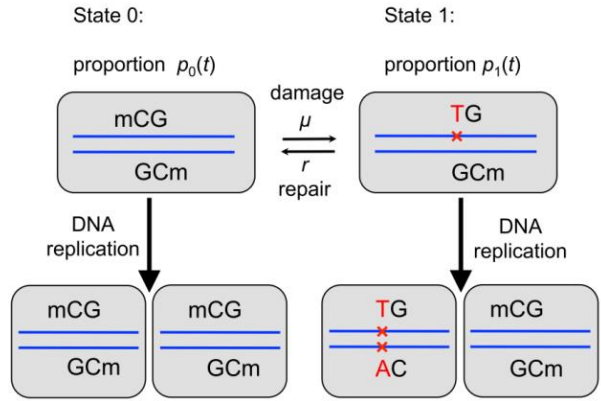

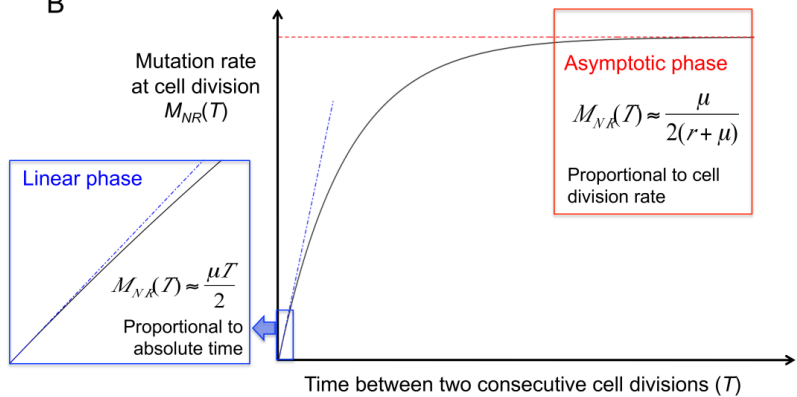

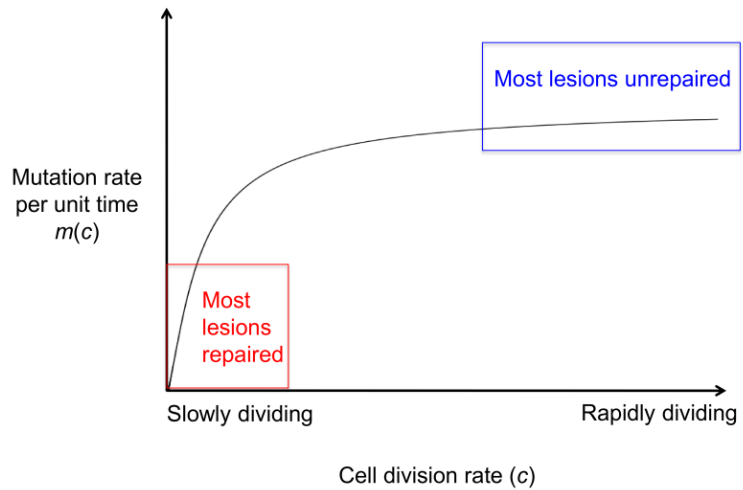

A) The DNA dynamics before and after replication. The upper panel shows the DNA states prior to the next cell division, and the lower panel shows the DNA states of the daughter cells after cell division.
B) The per cell division mutation rate increases with the time between two consecutive cell divisions and reaches an asymptote when the cell divides slowly.
C) The rate at which non-replicative mutations accumulate per unit time increases with the cell division rate.

**Figure S1. The effect of the generation time on the sex-averaged yearly rate of replication-driven mutations.**

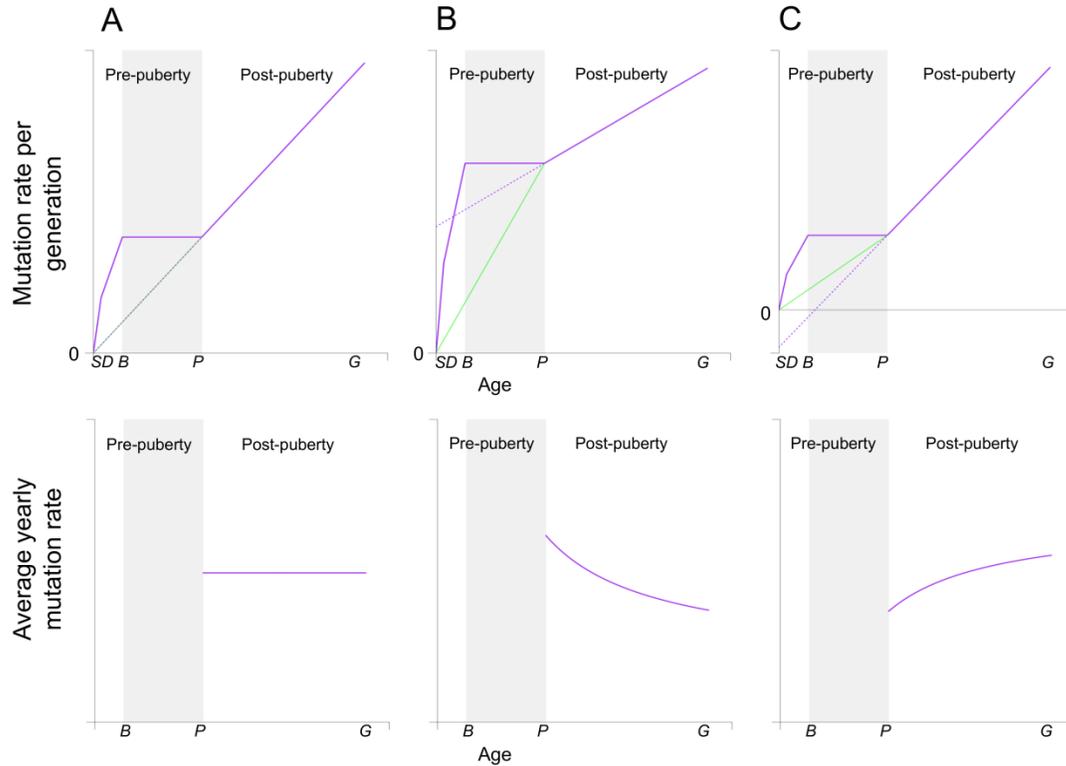

The sex-averaged mutation rate per generation (solid purple line) increases with the generation time, i.e., the reproductive age (assumed to be the same for males and females). Depending on the age of puberty (P), generation time (G), and the per cell division mutation rates, a linear fit to the number of mutations after puberty will have a zero, positive or negative intercept at age zero (i.e., fertilization). The green line connects the origin with the per generation mutation rate at puberty, so its slope represents the average yearly mutation rate prior to puberty. The slope of the dotted purple line represents the yearly mutation rate after puberty. The effect of G on the average yearly mutation rate depends on the relative slopes describing the accumulation of mutations pre and post puberty; its sign is equivalent to the sign of the intercept at age zero.

A) If the intercept is zero, the dotted purple and green lines coincide, and the yearly mutation rates before and after puberty are equal, so the generation time does not affect the average yearly mutation rate.
B) If the intercept is positive, the yearly mutation rate after puberty is smaller than that before puberty, so the average yearly mutation rate decreases with generation time (i.e., "the generation time effect").
C) If the intercept is negative, the yearly mutation rate after puberty is greater than that before puberty, so the average yearly mutation rate increases with generation time.

**Figure S2. A model for non-replicative mutations with errors in repair.**

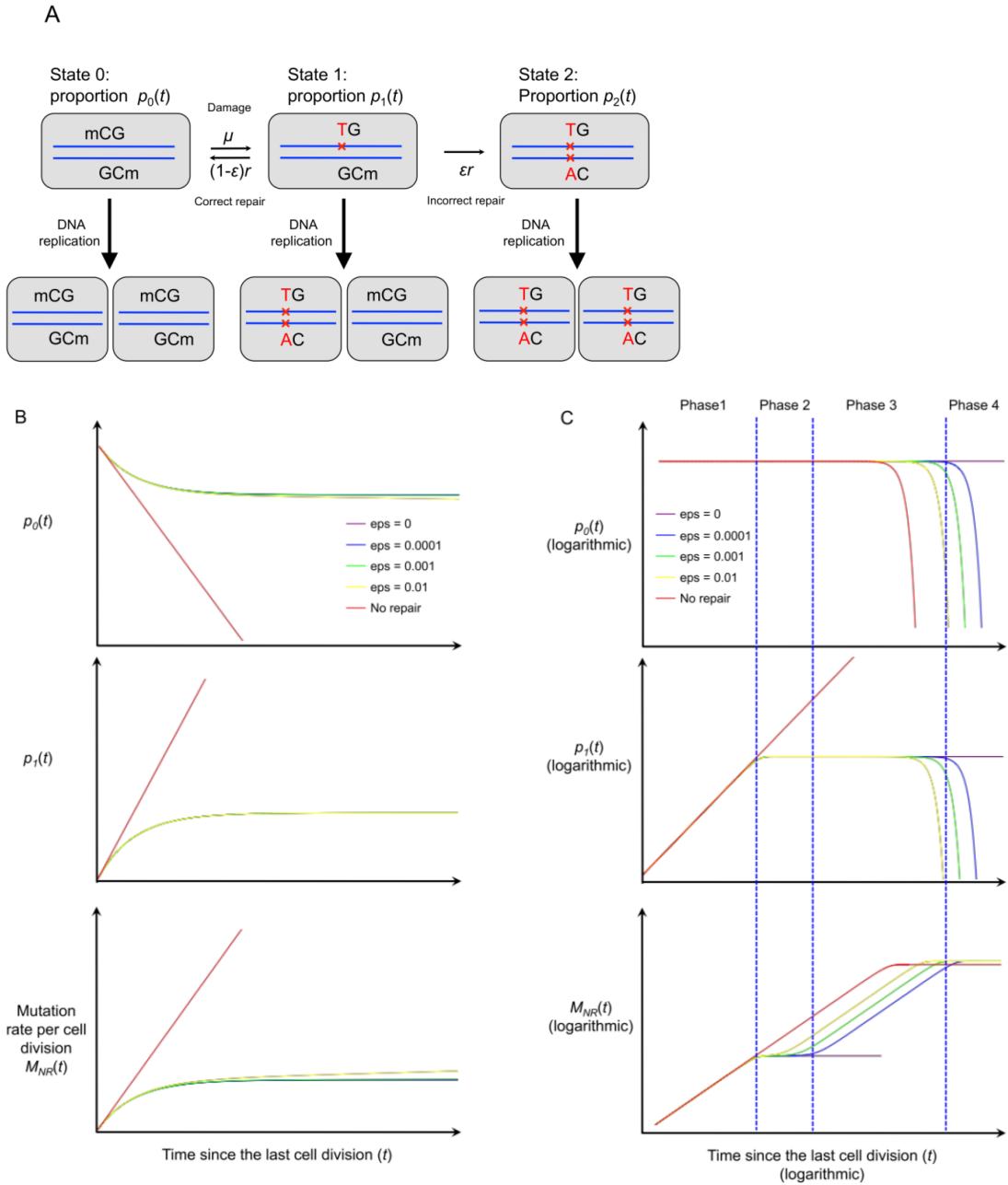

A) The DNA dynamics with errors in repair can be described by three states. The upper panel shows the DNA states prior to the next cell division, and the lower panel shows the DNA states of the daughter cells after cell division.

B) The proportion of base pairs without lesion ($p_0(t)$), the proportion of base pairs with single-strand lesions ($p_1(t)$) and the mutation rate per cell division ($M_{NR}(t)$) as functions of the time since the last division. Same values of the damage and repair rates are used for all cases with repair. In the case with no DNA repair, the value of $r$ is set to zero.

C) Log-log plots for $p_0(t)$, $p_1(t)$ and $M_{NR}(t)$. The dotted blue line show the boundaries between the four phases for the case with $\varepsilon$ =0.0001 (represented by the blue curve). Notice that both axes are on logarithmic scale, so later phases should be longer than they appear on the plot.

**Table S1. A list of parameters used in the model for replication-driven mutations.**

| Symbol | Definition | Estimate |
|---|---|---|
| $d_i^s$, $\mu_i^s$ | Number of cell divisions and replication error rate per division in the $i^{th}$ stage ($i$ =0, 1, 2, 3, 4) in sex $s$ ($s \in \{f, m\}$). Stage 0: the first post-zygotic division; Stage 1: from the second post-zygotic division to sex differentiation; Stage 2: from sex differentiation to birth; Stage 3: from birth to puberty; Stage 4: from puberty to reproduction. | $d_1 = 15$; $d_2^m = 21$; $d_2^f = 15$; $d_3^m = 0$; $d_3^f = 0$; $d_4^f = 0$; $d_4^m = c^m(G-P-t_{sg}) + d_{sg}$ |
| $t_{sg}$ | Duration of spermatogenesis (in years) | $t_{sg} = 0.2$ |
| $d_{sg}$ | number of cell divisions required to complete spermatogenesis from spermatogonial stem cells | $d_{sg} = 4$ |
| $c^m$ | Number of cell divisions undergone by spermatogonial stem cells in each year | $c^m = 23$ |
| $P$ | Age of puberty (assumed to be the same for both sexes) | $P = 13$ |
| $G$ | Age of reproduction (assumed to be the same for both sexes) | |
| $H$ | Total number of base pairs in a haploid set of autosomes | |
| $M_R^s$ | Numbers of autosomal replication-driven mutations inherited from the parent of sex $s$ | |
| $M_R$ | Total number of autosomal replication-driven mutations inherited by an offspring from both parents | |
| $m_{R,g}$ | Per generation mutation rate for replication-driven mutations | |
| $\alpha_R$ | Ratio of male to female replication-driven mutations | |
| $m_{R,y}$ | Average yearly mutation rate for replication-driven mutations | |

See Materials and methods for references behind each parameter value.

**Table S2. A list of parameters used in the model for non-replicative mutations.**

| Symbol | Definition |
|---|---|
| $\mu$ | Instantaneous damage rate |
| $R$ | Instantaneous repair rate |
| $R=\mu/r$ | Relative repair rate compared to damage rate |
| $p_0(t)$ | Proportion of base pairs in the genome that do not carry a lesion at time $t$ since last cell division |
| $p_1(t)$ | Proportion of base pairs in the genome that carry a single-strand lesion at time $t$ since last cell division |
| $p_2(t)$ | Proportion of base pairs in the genome that substituted at time $t$ since last cell division |
| $T$ | Time between two consecutive divisions of a cell lineage |
| $M_{NR}(T) = 0.5\, p_1(t) + p_2(t)$ | Mutation rate per division for a cell that divides every $T$ unit of time |
| $c=1/T$ | Cell division rate |
| $m(c) = c*M_{NR}(1/c)$ | Mutation rate per unit time for a cell with cell division rate $c$ |
| $\varepsilon$ | Error rate of repair |